\documentclass[prl,amsmath,amssymb,twocolumn,superscriptaddress, pdftex]{revtex4}
\usepackage[usenames]{color}
\usepackage{amssymb}
\usepackage{grffile}
\usepackage{amstext, amsfonts,amsxtra}
\usepackage{textcomp}
\usepackage{xspace}
\usepackage{color}

\begin{document}
\title{Cluster dynamics in two-dimensional lattice gases with inter-site interactions}
\author{Wei-Han Li}
\affiliation{Institut f\"ur Theoretische Physik, Leibniz Universit\"at Hannover, Appelstr. 2, 30167 Hannover, Germany}
\author{Arya Dhar}
\affiliation{Institut f\"ur Theoretische Physik, Leibniz Universit\"at Hannover, Appelstr. 2, 30167 Hannover, Germany}
\author{Xiaolong Deng}
\affiliation{Institut f\"ur Theoretische Physik, Leibniz Universit\"at Hannover, Appelstr. 2, 30167 Hannover, Germany}
\author{Luis Santos}
\affiliation{Institut f\"ur Theoretische Physik, Leibniz Universit\"at Hannover, Appelstr. 2, 30167 Hannover, Germany}

\begin{abstract}
Sufficiently strong inter-site interactions in extended-Hubbard and XXZ spin models result in dynamically-bound clusters at neighboring sites. 
We show that the dynamics of these clusters in two-dimensional lattices is remarkably different and richer than that of repulsively-bound on-site clusters in gases without inter-site interactions. 
Whereas on-site pairs move in the same lattice as individual particles, nearest-neighbor dimers perform an interacting quantum walk in a different lattice geometry, leading to 
a peculiar dynamics characterized by more than one time scale. The latter is general for any lattice geometry, but it 
is especially relevant in triangular and diamond lattices, where dimers move resonantly in an effective kagome and Lieb lattice, respectively. 
As a result, dimers experience partial quasi-localization due to an effective flat band, and may move slower than longer clusters. 
This surprising link between anomalously slow quantum walk dynamics in these models and flat-band physics 
may be readily observed in experiments with lanthanide atoms.
\end{abstract}
\maketitle



Experiments on atoms in optical lattices~\cite{Bloch2008} and trapped ions~\cite{Blatt2012} have revealed an intriguing dynamics resulting from the interplay between inter-particle interactions, 
hopping, and disorder~\cite{Billy2008,Roati2008,Deissler2010,Kondov2011,Jendrzejewski2012,Schreiber2015}, which has to a large extent motivated the interest on out-of-equilibrium 
quantum many-body systems in recent years~\cite{Polkovnikov2011,DAlessio2016,Borgonovi2016,Basko2006,Nandkishore2015}. 
Crucially, due to the almost perfect isolation that characterizes these experiments, lattice dynamics is to a large extent constrained by energy conservation. The 
latter may restrict the system to a particular manifold of states that depends on the initial preparation, and may lead to the formation of very long-lived metastable states.
This is best illustrated by repulsively-bound pairs~(RBPs), on-site couples of particles that, although thermodynamically unstable,  remain 
dynamically bound if the interaction strength exceeds the lattice bandwidth~\cite{Winkler2006,Strohmaier2010}, strongly handicapping particle motion~\cite{Schneider2012,Ronzheimer2013}. 

Constrained lattice dynamics becomes more intriguing in the presence of inter-site interactions. Although the latter may result from superexchange processses in short-range 
interacting systems, they are particularly relevant in a new generation of experiments 
with power-law interacting systems, including trapped ions~\cite{Richerme2014, Jurcevic2014}, Rydberg atoms~\cite{Browaeys2016}, and lattice gases with strong 
dipole-dipole interactions, in particular magnetic atoms and polar molecules. Seminal experiments on polar lattice gases have revealed 
inter-site spin-exchange~\cite{DePaz2013, Yan2013}, and realized an extended Hubbard model~(EHM) with nearest-neighbor interactions~\cite{Baier2016}.
The combination of energy conservation, finite bandwidth, and inter-site interactions is predicted to lead to dynamically-bound inter-site pairs and clusters~\cite{Valiente2009,Nguenang2009}, 
and even self-bound lattice droplets~\cite{Petrosyan2007,Li2020,Morera2020}. Experiments based on superexchange coupling have realized a one-dimensional 
interacting quantum walk of neighboring magnons in an XXZ spin chain~\cite{Fukuhara2013}. Nearest neighbor dimers~(NNDs) may significantly slow-down the dynamics in 1D EHMs~\cite{Barbiero2015}, and may induce
many-body quasi-localization in disorderless 1D polar lattice gases~\cite{Li2020}. 

In this Letter, we show that the dynamics of inter-site clusters in 2D lattices is strikingly different than that of on-site clusters in the absence of inter-site interactions. 
In 1D systems both on-site RBPs and NNDs move in second order in the same lattice as that of unpaired particles~(henceforth called singlons)~\cite{Winkler2006,Strohmaier2010,Valiente2009,Nguenang2009, Fukuhara2013, Barbiero2015,Li2020}. 
For on-site RBPs this remains true in any dimension. 
In stark contrast, we show that a bound NND perform an interacting quantum walk~\cite{Lahini2012} in an effective lattice with a different geometry than that of the singlon lattice, resulting in a peculiar 
dynamics, characterized by multiple time scales. Although the effect is general to all 2D geometries, it is particularly interesting in triangular and diamond lattices. 
In those lattices, although NNDs can move resonantly, dimer quasi-localization occurs due to the effective dimer lattice, which has a kagome and Lieb geometry, respectively, 
and hence presents a flat band~\cite{Leykam2018}. Although we focus below on polar gases, a similar intriguing dynamics of 
dimers and even longer inter-site clusters may be observed in other EHMs, as well as in 
magnon bound states in 2D XXZ chains.




\begin{figure}[t]
\begin{center}
\includegraphics[width =0.8\columnwidth]{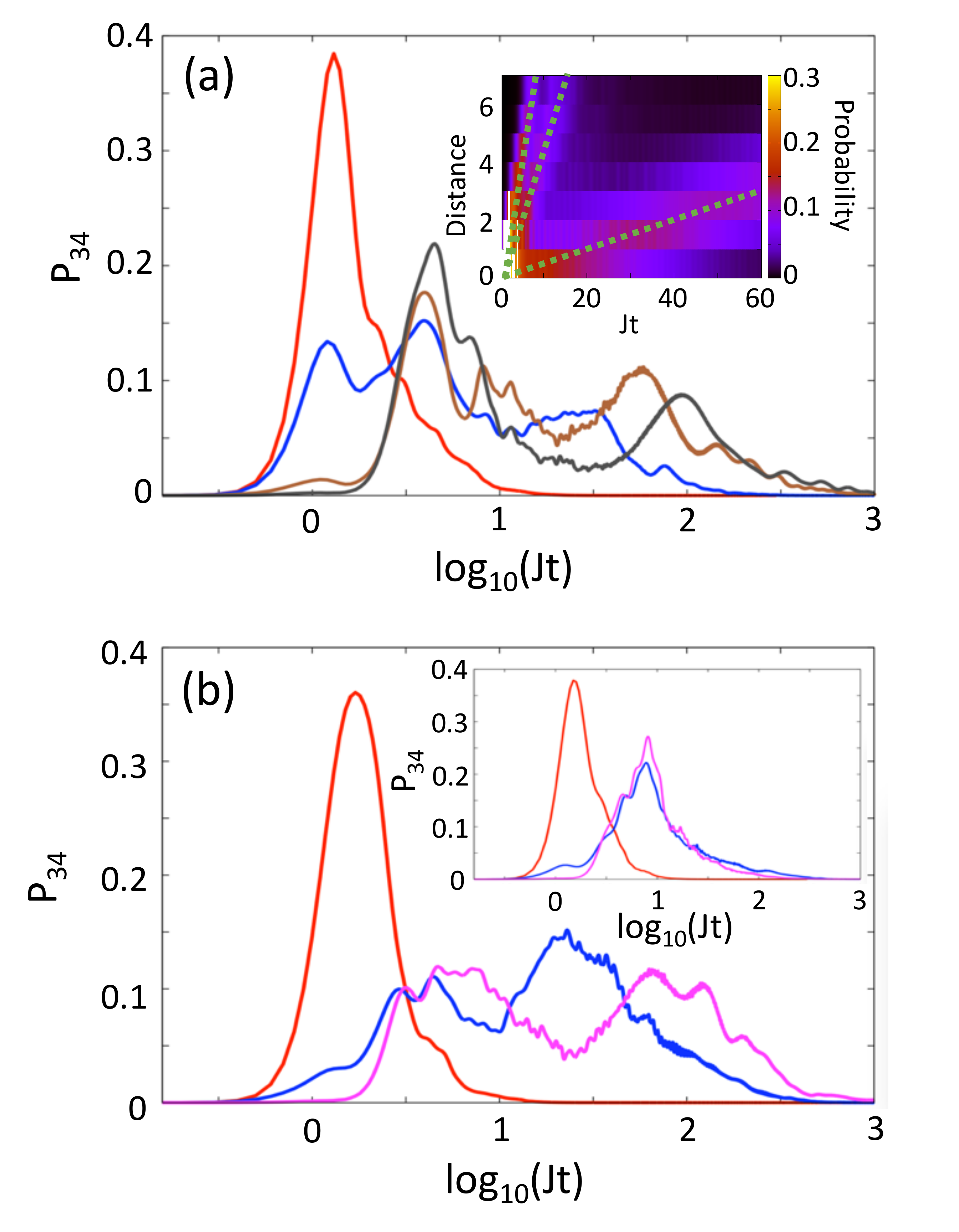}
\caption{Two-particle evolution. (a) $P_{34}$~(see text) for $V/J=0$~(red), $5$~(blue), $15$~(brown), and $30$~(grey).
Inset: probability for $V=20J$ of finding the center-of-mass at a given distance in lattice units from its initial position.
 Dotted lines are a guide to the eye. 
 (b) $P_{34}$ for $V/J=0$~(red), $15$~(blue), and $30$~(pink) for $J_H=0$. Inset: same for $J_H=0.5J$.  Results are obtained  
by exact evolution of Eq.~\eqref{eq:HT}~\cite{footnote-absorbing}, starting with two particles at NN.}
\label{fig:1}
\end{center}
\end{figure}




\paragraph{Triangular lattices.--}We consider a polar gas in a triangular lattice, with the dipoles orthogonal to the lattice plane. The 
system is well described by the Hamiltonian 
\begin{equation}
\hat H=-\sum_{\vec{j},\vec{s}} J_{\vec s} \left (  \hat b_{\vec{j}}^\dag \hat b_{ \vec{j}+\vec{s} } + \mathrm{H. c.} \right ) + 
\frac{V}{2}\sum_{\vec{i}\neq \vec{j}}\frac{1}{|\vec{r}_{\vec{i}}-\vec{r}_{\vec{j}}|^3} \hat n_{\vec{i}} \hat n_{\vec{j}}
\label{eq:HT}
\end{equation}
where $V$ characterizes the dipole-dipole interaction, $J_{\vec s}$ denotes the hopping rate to NN, $\vec{s}=\{(1,0), (0,1), (1,-1)\}$ and  $\hat n_{\vec{j}}=\hat b_{\vec{j}}^\dag \hat b_{\vec{j}}$, 
with $\hat b_{\vec{j}}$ the bosonic operator at site $\vec{j}=(j_1,j_2)$ placed at $\vec{r}_{\vec{j}}=j_1\vec{a}_1+j_2\vec{a}_2$, 
with $\vec{a}_1=\vec{e}_x$ and $\vec{a}_2=\vec{e}_x/2 + \sqrt{3} \vec{e}_y/2$. We assume a hard-core constraint $(\hat b_{\vec j}^\dag)^2=0$~(assuming a sufficiently large on-site interaction), 
and $J_{\vec s}=J$ for all $\vec s$~(we discuss below 
the case of different $J_{\vec s}$). 




\paragraph{Dynamically-bound dimers.--} Despite not being thermodynamically stable for $V>0$, NNDs, with energy $\simeq V$,  
remain dynamically stable, for strong-enough $V/J\gtrsim 10$~\cite{footnote-SM}. 
Whereas in 1D lattices, or in any other 2D lattice, NNDs move in second order with a hopping $\propto J^2/V$, dimer motion in triangular lattices is resonant, with hopping $J$, since  
a particle in the dimer can move to a neighboring site while still keeping a constant distance from its partner.
One could hence expect that NNDs move fast, in a time scale independent of $V$ for sufficiently large $V/J$. This expectation turns out to be incorrect. 

Figure~\ref{fig:1}(a) depicts for different $V/J$ as a function of time the probability $P_{34}$ of finding a particle between a distance of $3$ and $4$ lattice units 
from the initial center-of-mass of the dimer, obtained by exact evolution of Eq.~\eqref{eq:HT}~\cite{footnote-absorbing}~(the 
inset shows for $V/J=20$ the time dependence of the distance between the center-of-mass of the two-particle system and its initial position). 
Whereas for $V=0$ the particles expand ballistically with a velocity $\propto J$, when $V/J$ grows two markedly different expansion velocities become apparent, a fast one proportional to $J$, and a slow one 
proportional to $J^2/V$~\cite{footnote-SM}. These two time scales are apparent even down to $V/J=5$, at which dimers are still largely unbound~\cite{footnote-SM}.




\begin{figure}[t]
\begin{center}
\includegraphics[width =0.9\columnwidth]{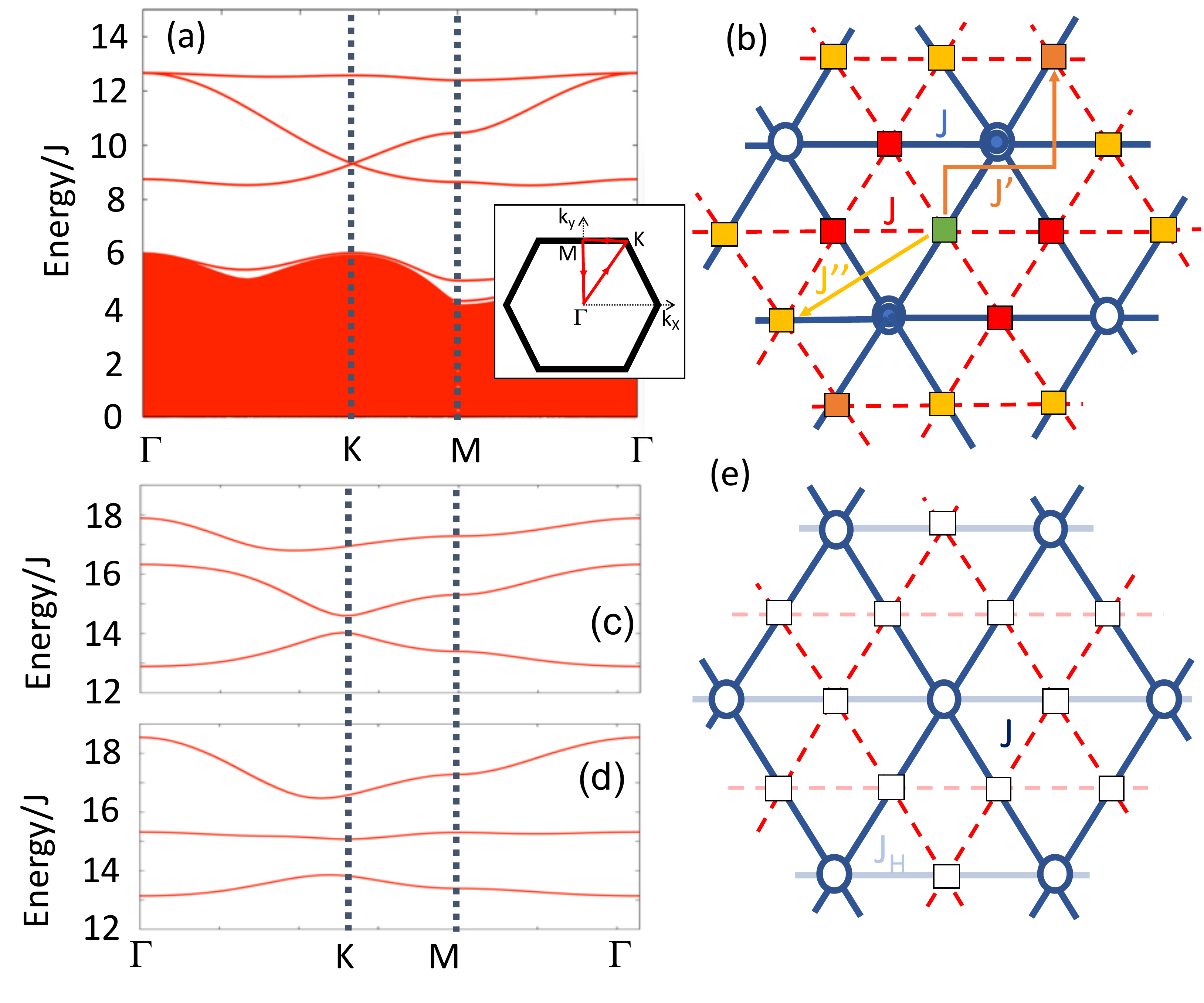}
\caption{(a) Two-particle eigenenergies in a triangular lattice for $V/J=10$ as a function of the center of mass quasi-momentum within the Brillouin zone~(inset). 
(b)  For large $V/J$, NNDs in a triangular lattice (solid lines, circular sites) move resonantly in a kagome lattice (dashed lines, square sites), with hopping $J$. 
A dimer initially at the green site, moves in first order into the red sites, and in second order into the orange/yellow ones~(see text). 
(c) and (d) bound dimer eigenstates for $V/J=15$ for, respectively, $J_H=0.5J$ and $J_H=0$. (e) Lattice with reduced hopping $J_H<J$ in one direction. 
The effective dimer lattice presents a flat band only for $J_H/J=\pm 1$~(kagome lattice) and  $0$~(Lieb lattice).}
\label{fig:2}
\end{center}
\end{figure}




\paragraph{Dimer bands and dimer lattice.--}The peculiar dimer dynamics stems from the actual dimer dispersion. 
For a given center of mass quasi-momentum $\vec K$, we calculate the two-particle spectrum associated to the relative coordinate 
between the particles, $E_\nu(\vec K)$~(Fig.~\ref{fig:2}(a))~\cite{footnote-SM}. As for RBPs, above a continuum of scattering states~(red continuum in Fig.~\ref{fig:2}(a)), 
we observe bound states characterized 
by a localized relative coordinate, and hence by sharp lines as a function of $\vec K$.
These bound states, with energy $\simeq V$, correspond to dynamically stable dimers, which for large $V/J$ are tightly bound at NN.

Whereas RBPs present a single band, polar dimers appear in three different bands, which 
are best understood in the large $V/J$ regime, of tightly-bound NNDs.
We associate to each NND an effective lattice site, at the middle point of the link joining the two particles (square sites in Fig.~\ref{fig:2}(b)). 
The motion of one of the particles to neighboring sites results in dimer hopping, with rate $J$, into one of four neighboring links~(red sites in Fig.~\ref{fig:2}(b)).  
Crucially, this resonant dimer motion does not take place in a triangular lattice, but in a kagome lattice with halved inter-site distance~(dashed lines in Fig.~\ref{fig:2}(b)). 
As it is well known~\cite{Leykam2018}, kagome lattices present three bands, including an upper flat band. 
The emerging kagome geometry hence results in the dimer bands observed in Fig.~\ref{fig:2}(a). 

The presence of the upper flat band in the dimer spectrum is responsible for the anomalously slow quantum walk dynamics observed in Fig.~\ref{fig:1}(a). 
An initially localized NND, projects into all three dimer bands with approximately equal probability. For large $V/J$, 
the dimer projection into the flat band~(which we denote as flat-band dimer~(FBD) henceforth) does not move in first order, and just expands due to second order processes, with hopping rates 
$J'=\frac{8}{7}\frac{J^2}{V}$ and $J''=\frac{3\sqrt{3}}{3\sqrt{3}-1}\frac{J^2}{V}$, that connect next-to-NN dimer sites in the effective kagome lattice~(see Fig.~\ref{fig:2}(b)). 
Additionally, the dispersive lower two dimer bands, which for the kagome lattice have an equal bandwidth, become clearly different at low $V/J$, with the lowest band markedly narrower than the 
middle one~\cite{footnote-SM}. As a result, for low $V/J$ the system presents an additional intermediate time scale, apparent for $V/J=15$ in Fig.~\ref{fig:1}(a)~(and in the inset of the figure). 
For growing $V/J$ the bandwidth of the lower band approaches that of the middle band, and hence, interestingly, the average velocity of non-FBD increases for growing $V/J$ until 
converging to a $V$ independent value.




\paragraph{Modified triangular lattices.--}
The crucial role played by the effective flat band becomes evident when considering the case of different hopping rates, e.g. $J_{(1,0)}=J_H<J$, $J_{(0,1)}=J_{(1,-1)}=J$~
(Fig.~\ref{fig:2}(e)). For $J_H=0.5 J$ there is no flat band and all bands have similar bandwidth. As a result, although the single-particle hopping is reduced, 
the dimer expansion dynamics presents a single time scale $\propto 1/J$~(inset of Fig.~\ref{fig:1}(b)). In contrast, for 
$J_H=0$, which corresponds to a diamond lattice, the NNDs experience an effective Lieb lattice, which is also characterized by the presence of a flat (middle) band~\cite{Leykam2018}~(see Fig.~\ref{fig:2}(d)). 
As a result, as for $J_H=J$,  a fast and a slow dimer dynamics are again clearly resolved~(Fig.~\ref{fig:1}(b)). 



\begin{figure}[t]
\begin{center}
\includegraphics[width =0.9\columnwidth]{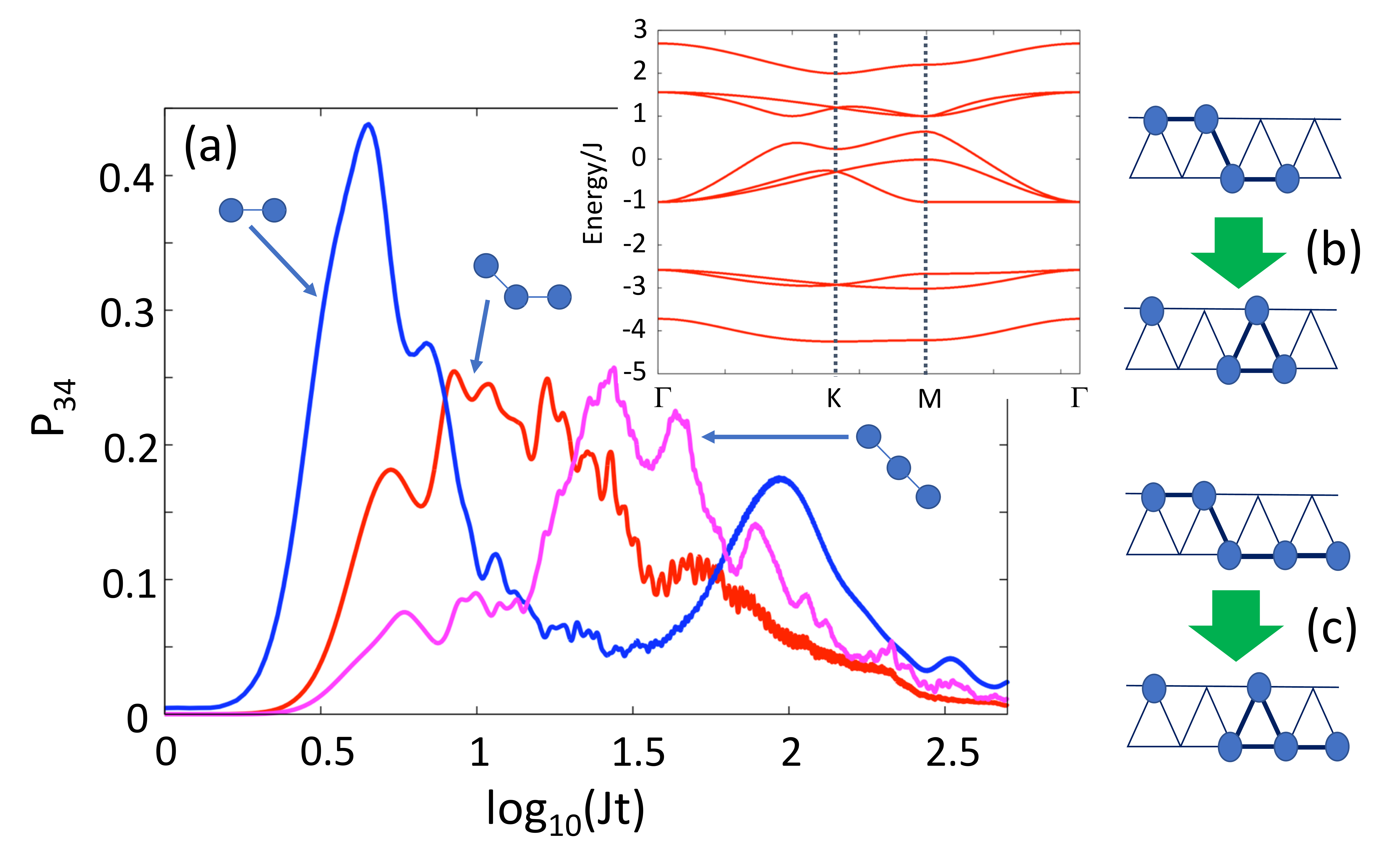}
\caption{Trimers and longer clusters. (a) $P_{34}(t)$ for $V=30J$ for a dimer~(blue) and two different initial tightly-bound trimer configurations~(red and pink). The initial trimers are indicated. 
Trimer results are obtained by exact evolution of the Hamiltonian for tightly-bound trimers~\cite{footnote-SM}. The inset depicts the energy bands of a linear trimer for $V=30J$. 
(b) Spontaneous singlon emission of a linear cluster with $4$~(b) and $5$~(c) particles, which creates a cluster with a filled elementary triangle that can only move 
in second order.}
\label{fig:3}
\end{center}
\end{figure}

\paragraph{Trimers and larger clusters.--} In stark contrast to gases without inter-site interactions, or gases with inter-site interactions in any other lattice geometry, inter-site-interacting gases in triangular lattices allow for the resonant motion of 
linear clusters of more than two particles, in which at most two sites are occupied in the same elementary triangle of the lattice.
A linear trimer has an energy $\simeq 2V$, and moves resonantly without spontaneously breaking into a dimer and a singlon. 
For a fixed position of the intermediate particle, there are $9$ possible trimer configurations, and hence tightly-bound trimers 
present $9$ energy bands~(inset of Fig.~\ref{fig:3}(a))~\cite{footnote-SM}.
None of them is flat, and hence trimers move in first order with a broad distribution of expansion velocities, but without revealing a bimodal expansion as for the dimer case~(Fig.~\ref{fig:3}(a)). 
Remarkably, since trimers move in first order, they move faster that FBDs for large-enough $V/J$ . 

Linear clusters with more than three particles decay spontaneously (resonantly) into a highly-mobile singlon and a cluster with at least one fully-occupied elementary triangle, 
which moves at most in second order~(see Figs.~\ref{fig:3}(b) and (c)).
Hence, dilute initially randomly-distributed particle distributions spontaneously decay into a gas of resonant movers~(singlons, non-FBDs, and linear trimers), 
and clusters that move at most in second-order~(FBDs and clusters with at least one fully occupied triangle).
We note at this point that hole clusters close to unit filling~(dilute holon gas) should behave similarly~\cite{footnote-holons}.


 
\begin{figure}[t]
\begin{center}
\includegraphics[width =0.9\columnwidth]{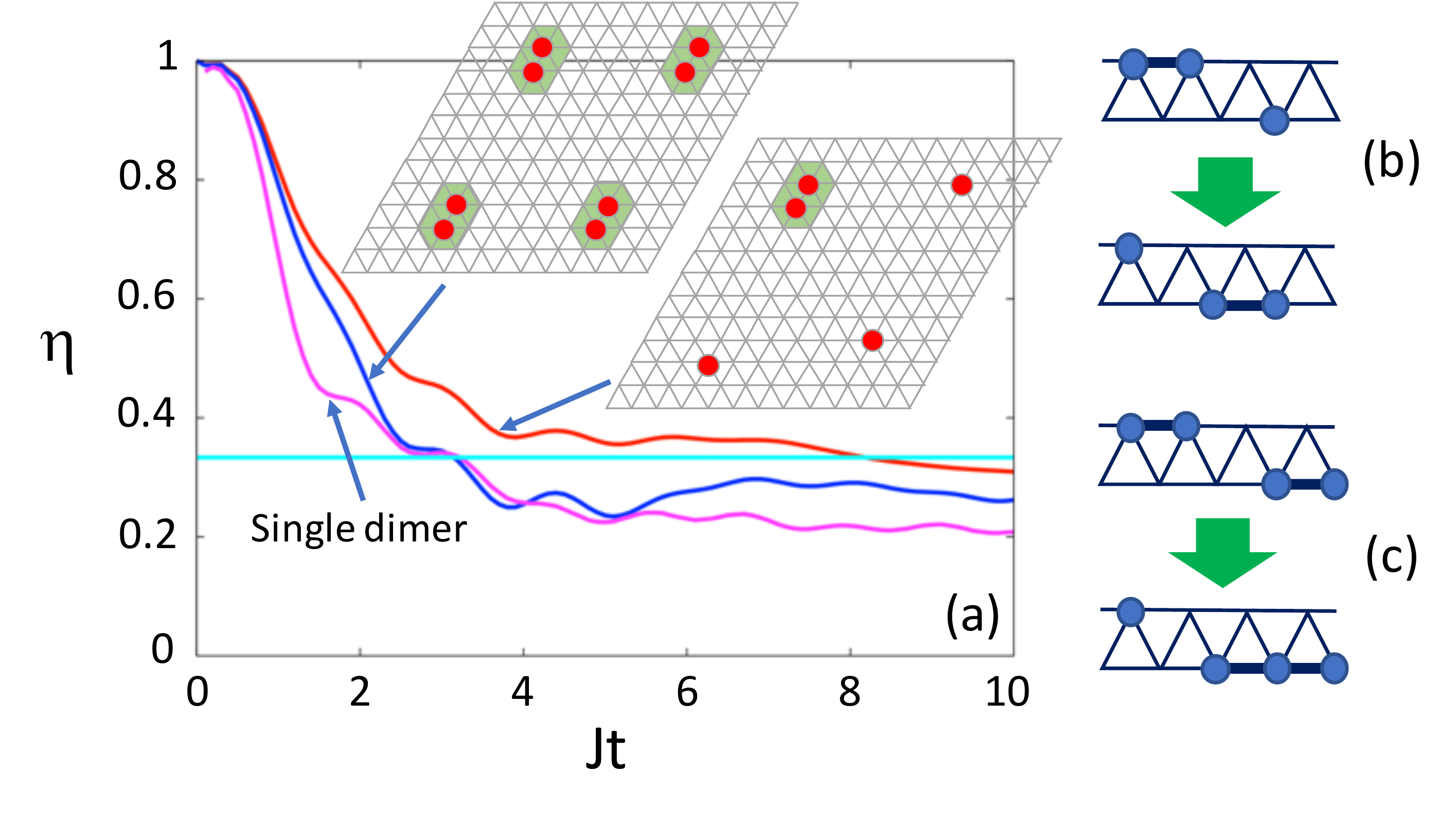}
\caption{(a) Homogeneity $\eta$~(see text) for $V=30J$ evaluated in a $13\times 13$ triangular lattice with open boundary conditions 
for a single dimer~(pink), a dimer and $3$ singlons~(red), and $4$ dimers~(blue). The insets indicate the initial conditions. 
The probability $P_s(t)$~(see text) is evaluated in the green regions in the insets. 
The light-blue line indicates the expected value $\eta=1/3$ for immobile FBDs.
The single-dimer results are obtained by exact evolution of Eq.~\eqref{eq:HT}. The other cases  were evaluated 
by means of TDVP calculations~\cite{footnote-TeNPy}, with a bond dimension up to $250$. 
(b) Singlon-dimer swap at nearest neighbors. (c) Dimer-dimer interactions may result  
in the resonant formation of a singlon and a linear trimer.}
\label{fig:4}
\end{center}
\end{figure}




\paragraph{Many-body dynamics.--} The presence of other particles or clusters may alter the cluster dynamics. 
Singlons and dimers at one site of distance may swap their positions inducing a singlon-mediated Brownian-like dimer motion~(Fig.~\ref{fig:4}(b)), that may also affect FBDs. 
Additionally,  two dimers may resonantly interchange a particle, resulting in a linear trimer and a singlon~(Fig.~\ref{fig:4}(b)). 
In polar lattice gases, these two processes are prevented for large-enough $V/J$ due to the blockade induced by the tail of the dipole-dipole interaction at next-NN. 
Moreover, if the clusters are not initially at one site of distance, resonant movers expand significantly before encountering non-resonant clusters, hence 
rendering many-body delocalization inefficient. For sparse fillings and large $V/J$, the main mechanism of FBD delocalization remains the 
second-order broadening of the flat band~($\propto J^2/V$). 

In Fig.~\ref{fig:4}(a) we show our results, obtained by means of time-dependent variational principle (TDVP) calculations~\cite{footnote-TeNPy}. 
We evaluate for a single dimer, one dimer and three singlons, and four dimers~(see the insets of Fig.~\ref{fig:4}(a)) the homogeneity 
$\eta = (P_s(t)-P_h)/(P_s(0)-P_h)$, where $P_s(t)$ is the average number of particles at the initial position of the dimers and their NN sites~(green regions in the insets of Fig.~\ref{fig:4}(a)), and $P_h$ is the expected 
value of $P_s$ if the dimers and singlons were homogeneously distributed in the lattice~\cite{footnote-SM}. $\eta=0$ would hence characterize an homogeneous distribution.
Due to the resonant movers, in all cases  $\eta$ decays within $t\lesssim 4/J$ to $\eta\simeq 1/3$, as expected for FBD quasi-localization~\cite{footnote-SM}.  
For a single dimer, $\eta$ plateaus, and only decays towards homogeneity ($\eta\to 0$) in a much longer time scale $\gg V/J^2$. The presence of other dimers 
should enhance FBD localization due to the  $1/r^3$ tail of the dipole-dipole interaction~(we are prevented to see this effect since our TDVP calculations are limited to 
$t\lesssim 20/J$). Indeed, since FBDs move with a hopping rate $\propto J^2/V$, as for the case of 1D dimers~\cite{Li2020} long-range dimer-dimer interactions are expected 
to lead to the clustering of FBDs, and hence FBD localization, if the mean distance between them 
is smaller than a critical value $\propto (V/J)^{2/3}$.



\begin{figure}[t]
\begin{center}
\includegraphics[width =0.9\columnwidth]{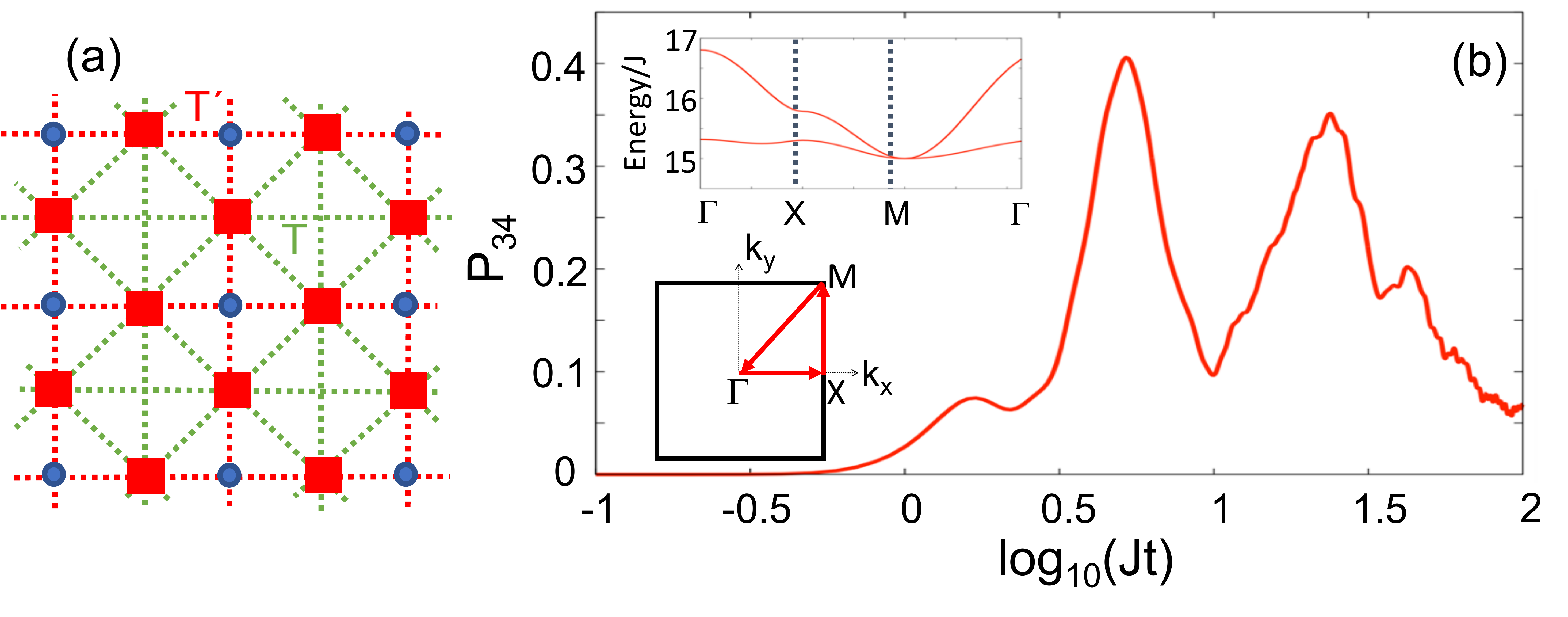}
\caption{Square lattice. (a) In a square lattice~(blue circles) NNDs move in second order in a decorated square lattice~(red squares), with hopping $T$~(green links), and $T'$~(red links), see text. (b) 
$P_{34}$ for $V=J=30$. The dimer expansion shows two distinct time scales. Results obtained  
by exact time evolution of Hamiltonian~\eqref{eq:HT}~\cite{footnote-absorbing}. 
In the upper inset we depict the bound dimer states as a function of the center-of-mass momentum in the first Brillouin zone~(lower inset).
}
\label{fig:5}
\end{center}
\end{figure}

\paragraph{Square lattice.--} The anomalous dimer dynamics resulting from an effective dimer lattice constitutes
a general feature for any 2D lattice geometry. This may be illustrated by a second relevant example, a square lattice, in which 
the hopping in Eq.~\eqref{eq:HT} becomes:
 $-J\sum_{\vec{j},\vec{s}} \left (  \hat b_{\vec{j}}^\dag \hat b_{ \vec{j}+\vec{s} } + \mathrm{H. c.} \right )$, with 
$\vec{s}=\{(1,0), (0,1))\}$. 
In contrast to the triangular lattice, dynamically-bound NNDs cannot move resonantly, but only in second order. 
As in the triangular lattice, the dimers move also in an effective lattice~(see Fig.~\ref{fig:5}(a)) that differs from the square lattice experienced by singlons. 
There are two different second-order dimer-hopping rates: 
$T=\frac{4\sqrt{2}}{2\sqrt{2}-1}\frac{J^2}{V}$~(green lines) 
and  $T'=\frac{8}{7}\frac{J^2}{V}$~(red lines).
These hops result in a decorated square lattice, with a four-site elementary cell, which is hence characterized by four dimer bands. 
In the inset of Fig.~\ref{fig:5}(b) we depict the two-particle spectrum in a square lattice as a function of the center of mass momentum $\vec K$. 
The two upper-bands correspond to the dimer states~(note that contrary to the triangular lattice, the Brillouin zone of the square lattice and that of the 
dimer lattice are not the same; as a result the four dimer bands map into the two bands observed in the two-particle spectrum).
Note that one of the bands is much narrower than the other, resulting, as for the triangular lattice, in two markedly different time scales~(Fig.~\ref{fig:5}(b)), although 
in a square lattice both of them are proportional to $V/J^2$. 
Finally, we would like to mention the case of honeycomb lattices, where dimer motion occurs also in second order, but in an effective kagome lattice~\cite{footnote-SM}. 
FBDs in honeycomb lattices are hence very strongly localized, being only able to move in fourth order~($\propto J^4/V^3$).



 
\paragraph{Conclusions.--} Two-dimensional polar lattice gases present an intriguing dynamics due to the interplay between energy conservation and 
inter-site interactions. Dynamically-bound dimers perform an interacting quantum walk in an effective lattice different than that of individual particles, resulting in 
an anomalous dynamics characterized by more than one time scale, which is particularly remarkable in triangular and diamond lattice due to the appearance of an emerging flat band.
Hence whereas individual particles and, remarkably, also trimers expand resonantly, 
dimer dynamics is, in contrast, largely handicapped by quantum interference. The required $V/J$ values are achievable with current state-of-the-art technology.   
For $^{164}$Dy in an UV lattice with $180$nm spacing and depth of $23$ recoil energies, $|V|/J\simeq 30$, with $J/\hbar\simeq 93$s$^{-1}$. 
Cluster dynamics in 2D lattices may then be probed in few seconds, well within the lifetimes in lanthanide experiments.
We stress, however, that, although the dipolar tail may play a relevant role in polar gases, the anomalous dimer dynamics just relies on 
nearest-neighbor interactions. Our results are 
hence generally valid for any 2D EHM, as well as for the dynamics of magnon bound states in 2D XXZ spin models.



\begin{acknowledgments}
We acknowledge support by the Deutsche Forschungsgemeinschaft (DFG, German Research Foundation) under the project SA 1031/11, 
the SFB 1227 ``DQ-mat'', project A04, and under Germany's Excellence Strategy -- EXC-2123 QuantumFrontiers -- 390837967.
\end{acknowledgments}






\begin{thebibliography}{99}

\bibitem{Bloch2008} I. Bloch, J. Dalibard, and W. Zwerger, Rev. Mod. Phys. {\bf 80}, 885 (2008).

\bibitem{Blatt2012} R. Blatt and C. F. Roos, Nature Phys. {\bf 8}, 277 (2012).

\bibitem{Billy2008} J. Billy, V. Josse, Z. Zuo, A. Bernard, B. Hambrecht, P. Lugan, D. Cl\'ement, L. Sanchez-Palencia, P. Bouyer, and A. Aspect, Nature {\bf 453}, 891 (2008).

\bibitem{Roati2008} 
G. Roati, C. D?Errico, L. Fallani, M. Fattori, C. Fort, M. Zaccanti, G. Modugno, M. Modugno, and M. Inguscio, Nature {\bf 453}, 895 (2008).

\bibitem{Deissler2010} B. Deissler, M. Zaccanti, G. Roati, C. D'Errico, M. Fattori, M. Modugno, G. Modugno, and M. Inguscio, Nat. Phys. {\bf 6}, 354 (2010).

\bibitem{Kondov2011} S. S. Kondov, W. R. McGehee, J. J. Zirbel, and B. DeMarco, Science {\bf 334}, 66 (2011).

\bibitem{Jendrzejewski2012} F. Jendrzejewski, A. Bernard, K. M\"{u}ller, P. Cheinet, V. Josse, M. Piraud, L. Pezz\'e, L. Sanchez-Palencia, A. Aspect, and P. Bouyer, Nat. Phys. {\bf 8}, 398 (2012).

\bibitem{Schreiber2015} 
M. Schreiber, S. S. Hodgman, P. Bordia, H. P. L\"uschen, M. H. Fischer, R. Vosk, E. Altman, U. Schneider, and I. Bloch, Science {\bf 349}, 842 (2015).

\bibitem{Polkovnikov2011} A. Polkovnikov, K. Sengupta, A. Silva, and M. Vengalattore, Rev. Mod. Phys. {\bf 83}, 863 (2011).

\bibitem{DAlessio2016} L. D'Alessio, Y. Kafri, A. Polkovnikov, and M. Rigol, Adv. Phys. {\bf 65}, 239 (2016).

\bibitem{Borgonovi2016} F. Borgonovi, F. M. Izrailev, L. F. Santos, and V. G. Zelevinsky, Phys. Rep. {\bf 626}, 1 (2016).

\bibitem{Basko2006} D. M. Basko, I. L. Aleiner, and B. L. Altshuler, Ann. Phys. {\bf 321}, 1126 (2006).

\bibitem{Nandkishore2015} R. Nandkishore and D. A. Huse, Annu. Rev. Condens. Matter Phys. {\bf 6}, 15 (2015). 

\bibitem{Winkler2006} K. Winkler, G. Thalhammer, F. Lang, R. Grimm, J. Hecker Denschlag, A. J. Daley, A. Kantian, H. P. B\"uchler, and P. Zoller, Nature {\bf 441}, 853 (2006).

\bibitem{Strohmaier2010} N. Strohmaier, D. Greif, R. J\"ordens, L. Tarruell, H. Moritz, T. Esslinger, R. Sensarma, D. Pekker, E. Altman, and E. Demler, Phys. Rev. Lett. {\bf 104}, 080401 (2010).

\bibitem{Schneider2012} 
U. Schneider, L. Hackerm\"uller, J. P. Ronzheimer, S. Will, S. Braun, T. Best, I. Bloch, E. Demler, S. Mandt, D. Rasch, and A. Rosch, Nature Phys. {\bf 8}, 213 (2012).

\bibitem{Ronzheimer2013} J. P. Ronzheimer, M. Schreiber, S. Braun, S. S. Hodgman, S. Langer, I. P. McCulloch, F. Heidrich-Meisner, I. Bloch, and U. Schneider, Phys. Rev. Lett. {\bf 110}, 205301 (2013).

\bibitem{Richerme2014} 
P. Richerme, Z.-X. Gong, A. Lee, C. Senko, J. Smith, M. Foss-Feig, S. Michalakis, A. V. Gorshkov, and C. Monroe, Nature {\bf 511}, 198 (2014).

\bibitem{Jurcevic2014} P. Jurcevic, B. P. Lanyon, P. Hauke, C. Hempel, P. Zoller, R. Blatt, and C. F. Roos, Nature {\bf 511}, 202 (2014).

\bibitem{Browaeys2016} A. Browaeys and  T. Lahaye (2016) Interacting Cold Rydberg Atoms: A Toy Many-Body System. In: O. Darrigol, B. Duplantier, J. M. Raimond, and V. Rivasseau (eds) 
Niels Bohr, 1913-2013. Progress in Mathematical Physics, vol 68. Birkh\"{a}user, Cham.

\bibitem{DePaz2013} A. de Paz, A. Sharma, A. Chotia, E. Mar\'{e}chal, J. H. Huckans, P. Pedri, L. Santos, O. Gorceix, L. Vernac, and B. Laburthe-Tolra, Phys. Rev. Lett. {\bf 111}, 185305 (2013).

\bibitem{Yan2013} B. Yan, S. A. Moses, B. Gadway, J. P. Covey, K. R. A. Hazzard, A. M. Rey, D. S. Jin, and J. Ye , Nature (London) {\bf 501}, 521 (2013).

\bibitem{Baier2016} 
S. Baier, M. J. Mark, D. Petter, K. Aikawa, L. Chomaz, Z. Cai, M. Baranov, P. Zoller, and F. Ferlaino, Science {\bf 352}, 201 (2016).

\bibitem{Valiente2009} M. Valiente and D. Petrosyan, J. Phys. B: At. Mol. Opt. Phys. {\bf 42}, 121001 (2009).

\bibitem{Nguenang2009} J.-P. Nguenang and S. Flach, Phys. Rev. A {\bf 80}, 015601 (2009).

\bibitem{Petrosyan2007} D. Petrosyan, B. Schmidt, J. R. Anglin, and M. Fleischhauer, Phys. Rev. A76, 033606 (2007).

\bibitem{Li2020} W. Li, A. Dhar, X. Deng, K. Kasamatsu, L. Barbiero, and L. Santos, Phys. Rev. Lett. {\bf 124}, 010404 (2020).

\bibitem{Morera2020} I. Morera, G. E. Astrakharchik, A. Polls, B. Juli\'{a}-D\'{i}az, arXiv:2007.01786.

\bibitem{Fukuhara2013}  T. Fukuhara, P. Schau\ss\ , M. Endres, S. Hild, M. Cheneau, I. Bloch, and C. Gross, Nature {\bf 502}, 76 (2013). 

\bibitem{Barbiero2015} L. Barbiero, C. Menotti, A. Recati, and L. Santos, Phys. Rev. B {\bf 92}, 180406 (2015).

\bibitem{Lahini2012} Y. Lahini, M. Verbin, S. D. Huber, Y. Bromberg, R. Pugatch, and Y. Silberberg, Phys. Rev. A {\bf 86}, 011603(R) (2012).

\bibitem{Leykam2018} D. Leykam, A. Andreanov, and S. Flach, Adv. Phys.: X {\bf 3}, 1473052 (2018)

\bibitem{footnote-absorbing} We consider a $13\times 13$ sites, along $\vec{a}_1$ and $\vec{a}_2$, imposing absorbing boundary conditions. The latter is necessary to observe 
the slow expansion of FBDs, without the distorting effect introduced by the reflection of fast-expanding dimers at the lattice boundaries.

\bibitem{footnote-SM} See the Supplementary Material for further details.

\bibitem{footnote-holons}  Note however that the EHM does not 
fulfill an exact particle-hole symmetry. In particular, holons are reflected at the lattice boundaries due to the different coordination number of edge sites.

\bibitem{footnote-TeNPy} The TDVP calculations were performed using the TeNPy Library~(version 0.5.0)~\cite{Hauschild2018}.

\bibitem{Hauschild2018} J. Hauschild and F. Pollmann, SciPost Phys. Lect. Notes {\bf 5} (2018).

\end{thebibliography}
\end{document}


\title{Supplementary Material for "Cluster dynamics in two-dimensional polar lattice gases"}
\author{Wei-Han Li}
\affiliation{Institut f\"ur Theoretische Physik, Leibniz Universit\"at Hannover, Appelstr. 2, 30167 Hannover, Germany}
\author{Arya Dhar}
\affiliation{Institut f\"ur Theoretische Physik, Leibniz Universit\"at Hannover, Appelstr. 2, 30167 Hannover, Germany}
\author{Xiaolong Deng}
\affiliation{Institut f\"ur Theoretische Physik, Leibniz Universit\"at Hannover, Appelstr. 2, 30167 Hannover, Germany}
\author{Luis Santos}
\affiliation{Institut f\"ur Theoretische Physik, Leibniz Universit\"at Hannover, Appelstr. 2, 30167 Hannover, Germany}

\maketitle

\paragraph{Two-particle eigenstates.--}  Two-particle states are parameterized by the particle positions, $|\vec r_A, \vec r_B\rangle$, 
where $\vec r_{j=A,B} = j_{1j}\vec a_1 + j_{2j} \vec a_2$.  
It is convenient to express the states in terms of the center-of-mass $\vec R=(\vec r_A+\vec r_B)/2$ and relative coordinate $\vec r = \vec r_A-\vec r_B$. 
The Hamiltonian for two particles acquires the form:
\begin{eqnarray}
H_D &=& -\sum_{\vec R, \vec r, \vec s} J_{\vec s} \left \{ \left [ |\vec R+\vec s/2, \vec r+\vec s\rangle \right\delimiter0 \right\delimiter 0  \nonumber \\
&& \left \delimiter 0 \left \delimiter 0 +   |\vec R+\vec s/2, \vec r-\vec s\rangle \right ]\langle \vec R, \vec r | + \mathrm{H. c.}  \right \}  \nonumber \\
&+& \sum_{\vec R,\vec r} \frac{V}{D(\vec r)^3} |\vec R,\vec r\rangle\langle \vec R,\vec r|.
\end{eqnarray}
where $D(\vec r=(j_1,j_2))=\sqrt{j_1^2+j_2^2+j_1j_2}$ is the distance between the particles. 
Fourier-transforming, we obtain the Hamiltonian $H=\sum_{\vec K} H_D(\vec K)$, with $\vec K$ the center-of-mass momentum, and 
\begin{eqnarray}
H_D(\vec K) &=& -\sum_{\vec s, \vec r}J_{\vec s}(\vec K) \left [ |\vec K,\vec r+\vec s\rangle\langle \vec K,\vec r| + \mathrm{H. c.} \right ]  \nonumber \\
&+& \sum_{\vec r} \frac{V}{D(\vec r)^3} |\vec K,\vec r\rangle\langle \vec K,\vec r|,
\end{eqnarray}
with $J_{\vec s}(\vec K)=2J_{\vec s}\cos\left (\frac{\vec K\cdot \vec s}{2}\right )$. When evaluating the previous Hamiltonian, we should avoid double counting, and hence 
we impose $\vec r = (j_1,j_2)$, such that $j_1\leq 0$, and for $j_1=0$, $j_2>0$. This requires particular care at $j_1=0$. In particular, we impose 
that the hop in the $(-1,0)$ direction couples $(1,j_2\le 0)$ with $(0,-j_2)$, and the hop in the $(-1,1)$ direction couples $(1,j_2<-1)$ with $(0,-(1+j_2))$. 
We diagonalize $H_D(\vec K)$ for each $\vec K$, which provides the eigenstates $E_\nu(\vec K)$ depicted in the main text. 
For the square lattice, the procedure is identical, except for the absence of the $J_{(1,-1)}$ hop, and that $D(\vec r=(j_x,j_y))=\sqrt{j_x^2+j_y^2}$.

\begin{figure}[t]
\begin{center}
\includegraphics[width =\columnwidth]{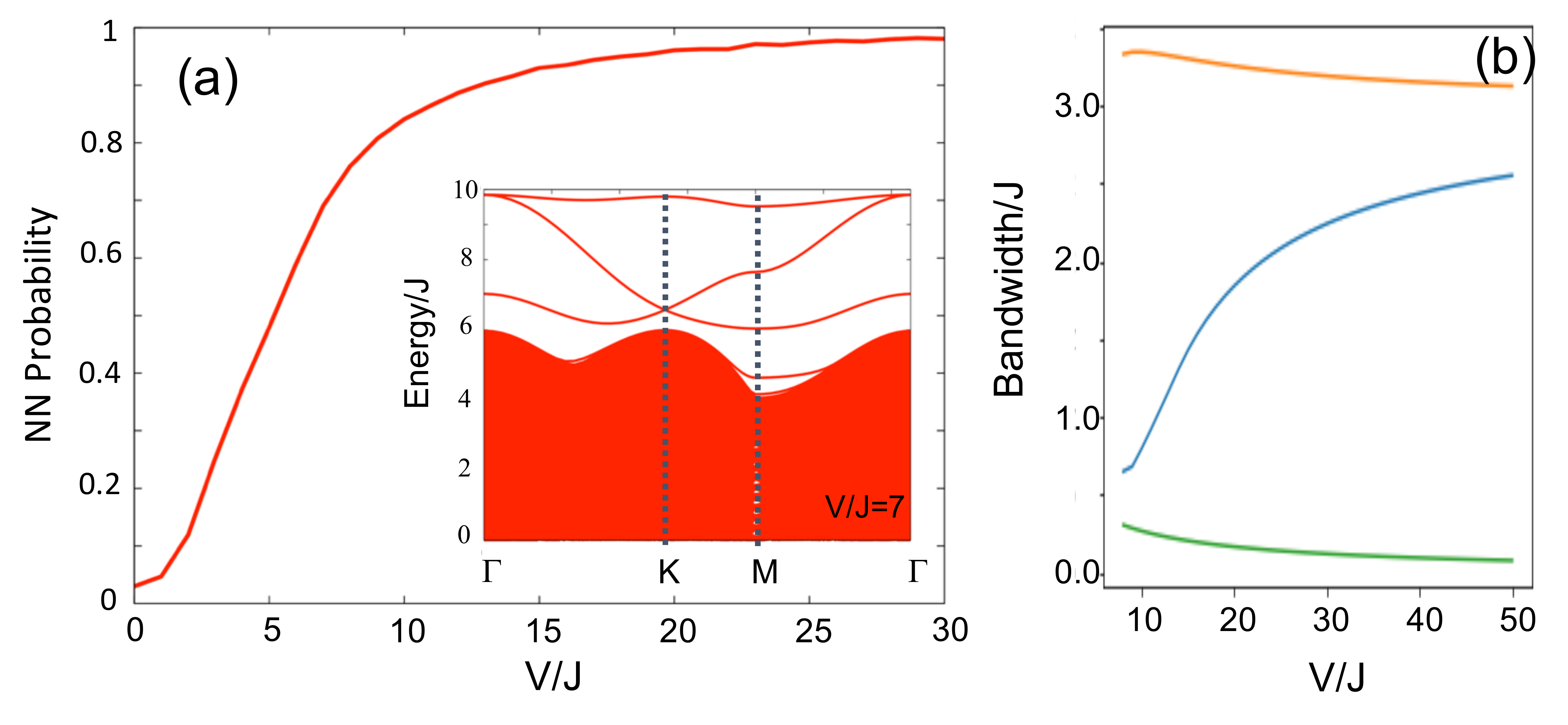}
\caption{Two-particle states. (a) Probability to find the two particles at nearest neighbors at $t=10/J$ for different $V/J$ values. Results obtained by exact time evolution of $H_D$ for two particles initially placed at nearest neighbors. In the inset we depict the two-particle eigenstates for $V/J=7$. Note that the three bound dimer states are already fully separated from the continuum of scattering states. 
(b) Bandwidth of the lower~(blue), middle~(yellow), and upper~(green) dimer bands.}
\label{fig:S1}
\end{center}
\end{figure}

For $V/J=7$~(see the inset of Fig.~\ref{fig:S1}(a)) the bound dimer states are already fully separated from the scattering states. For $V/J\gtrsim 10$ nearest-neighbor dimers remain tightly bound, as shown in 
Fig.~\ref{fig:S1}(a) where we evaluate by means of exact evolution of Hamiltonian $H_D$ 
 the probability that two particles initially at nearest neighbors remain still at nearest neighbors for $t=10/J$ . 
 
 As mentioned in the main text, the dimer bands approach the bands of a kagome lattice for large $V/J$. However, for intermediate $V/J$ values, although the dimers 
 are already bound, the bands are significantly distorted compared to the exact kagome case. Figure~\ref{fig:S1}(b) depicts the bandwidth for 
 the lower~(blue), middle~(yellow), and upper~(green) dimer bands. Note that the upper band is very narrow even at low $V/J$, and becomes even narrower for larger $V/J$~(as discussed 
 in the main text its band width decays as $J^2/V$). Note also that whereas in the kagome lattice the two dispersive bands have equal width, for low $V/J$ the lowest band is significantly narrower. 
 This may lead to an observable third expansion velocity as for $V/J=15$ in Fig. 1(a) of the main text (and the inset of that figure). Interestingly, this also means that increasing $V/J$ 
 the average velocity of non-flat-band dimers increases up to a $V$-independent value for large $V/J$.

\begin{figure}[t]
\begin{center}
\includegraphics[width =0.8\columnwidth]{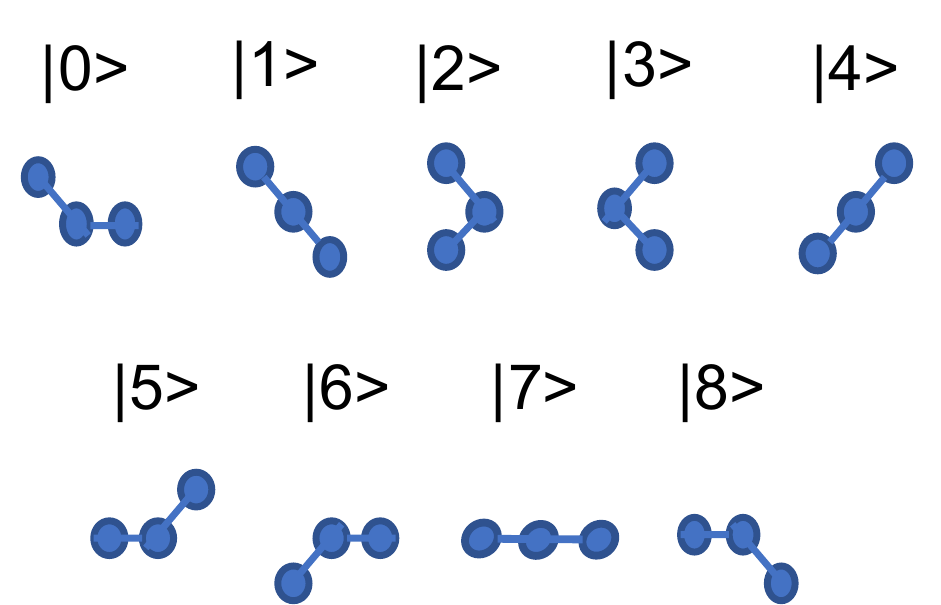}
\caption{Possible trimer configurations for a fixed central particle.}
\label{fig:S2}
\end{center}
\end{figure}

\paragraph{Linear trimers.--} Tightly-bound linear trimer states $|\vec n, \vec r\rangle$ are characterized by the position $\vec r$ of the central particle, 
and by the configuration $n=0,\dots,8$ of the other two particles with respect to the central one (see Fig.~\ref{fig:S2}). 
The Hamiltonian for a linear trimer is of the form: $H_T=H_{T0}+H_{T1}+H_{TI}$, where 
\begin{eqnarray}
H_{T0} &=&-J\sum_{\vec r} \left [ \left ( |0,\vec r\rangle + |2,\vec r\rangle + |3,\vec r\rangle + |7,\vec r\rangle \right ) \langle 1,\vec r|   \right \delimiter 0 \nonumber \\
&& + \left ( |2,\vec r\rangle + |3,\vec r\rangle + |5 ,\vec r\rangle + |6,\vec r\rangle \right) \langle 4,\vec r|  \nonumber \\
&& \left\delimiter 0 +\left ( |0,\vec r\rangle + |5,\vec r\rangle + |6 ,\vec r\rangle + |8,\vec r\rangle \right) \langle 7,\vec r| + \mathrm{H. c.}  \right ] \nonumber
\end{eqnarray}
is given by the hops of the side particles without changing the position of the central one, 
\begin{eqnarray}
H_{T1}&=&-J\sum_{\vec r} \left [ |8,\vec r+(0,1)\rangle\langle 0, \vec r| \right \delimiter 0 \nonumber \\
&& +  |2,\vec r+(0,1)\rangle\langle 3, \vec r| \nonumber \\
&& \left\delimiter 0 + |5,\vec r+(1,-1)\rangle\langle 6, \vec r| + \mathrm{H. c.} \right ] \nonumber 
\end{eqnarray}
contains the hops of the central particle in the triangular lattice, and 
\begin{eqnarray}
H_{TI}&=& \left ( \frac{V}{8} - \frac{V}{3\sqrt{3}}\right ) \left [
|1,\vec r\rangle\langle 1,\vec r|  \right \delimiter 0 \nonumber \\
&&\left\delimiter 0  + |4,\vec r\rangle\langle 4,\vec r| + |7,\vec r\rangle\langle 7,\vec r| \right ]
\end{eqnarray}
is the interaction energy difference, given by next-to-nearest neighbor dipole-dipole interactions, between 
states $|1\rangle$, $|4\rangle$, and $|7\rangle$, and the rest. The trimer evolution depicted in Fig. 4 of the main text is obtained 
by exact evolution of $H_T$.

\begin{figure}[t]
\begin{center}
\includegraphics[width =0.6\columnwidth]{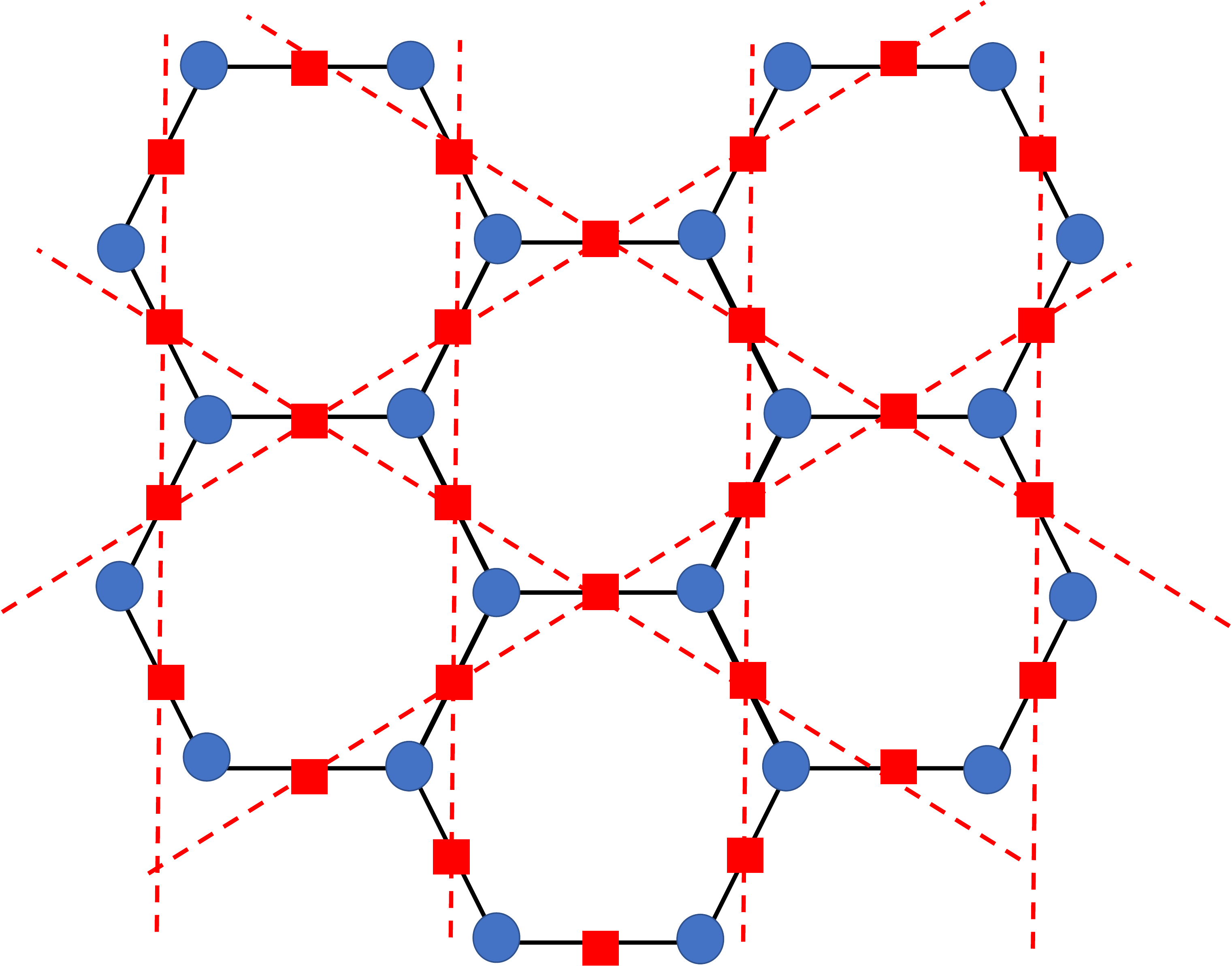}
\caption{Effective dimer lattice~(red squares and dashed lines) for a honeycomb lattice~(blue circles and solid lines). The dimer lattice has a kagome geometry.}
\label{fig:S3}
\end{center}
\end{figure}

As for the dimer states, we Fourier-transform $|n,\vec r\rangle \to |n,\vec k\rangle$. The Hamiltonian can be then written as 
$H_T=\sum_{\vec k} H_T(\vec k)$, where $H_T(\vec k)=H_{T0}(\vec k)+H_{T1}(\vec k)+H_{TI}(\vec k)$, with 
\begin{eqnarray}
H_{T0}(\vec k) &=&-J\left [ \left ( |0,\vec k\rangle + |2,\vec k\rangle + |3,\vec k\rangle + |7,\vec k\rangle \right ) \langle 1,\vec k|   \right \delimiter 0 \nonumber \\
&& + \left ( |2,\vec k\rangle + |3,\vec k\rangle + |5 ,\vec k\rangle + |6,\vec k\rangle \right) \langle 4,\vec k|  \nonumber \\
&& \left\delimiter 0 +\left ( |0,\vec k\rangle + |5,\vec k\rangle + |6 ,\vec k\rangle + |8,\vec k\rangle \right) \langle 7,\vec k| + \mathrm{H. c.}  \right ] \nonumber 
\end{eqnarray}
\begin{eqnarray}
H_{T1}(\vec k)&=&-J\sum_{\vec r} \left [ e^{ik_2} |8,\vec k\rangle\langle 0, \vec k| \right \delimiter 0 \nonumber \\
&& +  e^{ik_1} |2,\vec k\rangle\langle 3, \vec k| \nonumber \\
&& \left\delimiter 0 + e^{i(k_1-k_2)} |5,\vec k\rangle\langle 6, \vec k| + \mathrm{H. c.} \right ], \nonumber 
\end{eqnarray}
\begin{eqnarray}
H_{TI}(\vec k)&=& \left ( \frac{V}{8} - \frac{V}{3\sqrt{3}}\right ) \left [
|1,\vec k\rangle\langle 1,\vec k|  \right \delimiter 0 \nonumber \\
&&\left\delimiter 0  + |4,\vec k\rangle\langle 4,\vec k| + |7,\vec k\rangle\langle 7,\vec k| \right ]
\end{eqnarray}
We diagonalize for each quasi-momentum $\vec k$ within the first Brillouin zone obtaining the results presented in the main text.

\paragraph{Homogenization.--} We estimate the excluded volume as half of the number of NN 
of singlons and dimers~(induced by the NN blockade) plus the number of occupied sites~(due to the hard-core constraint). This results in the estimation
$P_h=\frac{10 N_D  N_P}{L^2 - 4 N_S - 6 N_D}$, where $N_S$ is the number of singlons, $N_D$ the number of dimers, $N_P=N_S+2N_D$ is the number of particles, 
$L^2$ is the number of lattice sites, and $10 N_D$ is the number of sites in the green regions of the insets of Fig.~4(a). Since each dimer projects $1/3$ into the flat band, 
we estimate $P_s$ for localized FBDs as $\frac{2}{3}P_h + \frac{1}{3} 2 N_D$, which results in $\eta=1/3$.

\paragraph{Honeycomb lattice.--} 
Fig.~\ref{fig:S3} depicts a honeycomb lattice, showing that the dimer lattice is, as for the triangular lattice, also a kagome lattice.